\documentclass[amsmath,amssymb,prr,superscriptaddress,reprint,showpacs,longbibliography]{revtex4-1}
\usepackage{graphicx}
\usepackage{dcolumn}
\usepackage[colorlinks=true,linkcolor=blue,citecolor=blue,urlcolor=blue]{hyperref}
\usepackage[usenames,dvipsnames]{xcolor}
\usepackage{soul}
\usepackage{braket}
\usepackage{bm}
\usepackage{upgreek}
\usepackage{mathtools}
\usepackage{bbm}

\renewcommand{\vec}[1]{{\bf #1}}

\begin{document}

\title{Magnetism and Magnetotransport in the Kagome Antiferromagnet $\text{Mn}_3\text{Ge}$}


\author{Gaurav Chaudhary}
\affiliation{Materials Science Division, Argonne National Laboratory, Lemont, IL 60439, USA}

\author{Anton A. Burkov}
\affiliation{Department of Physics and Astronomy, University of Waterloo, Waterloo, Ontario N2L 3G1, Canada}
\affiliation{Perimeter Institute for Theoretical Physics, Waterloo, Ontario N2L 2Y5, Canada}

\author{Olle G. Heinonen}
\affiliation{Materials Science Division, Argonne National Laboratory, Lemont, IL 60439, USA}

\date{\today}

\pacs{}
\keywords{}


\begin{abstract}
We perform classical Monte Carlo and stochastic Landau-Lifshitz-Gilbert simulations to study temperature dependent magnetism of Kagome antiferromagnet (AFM) Weyl metal $\text{Mn}_3\text{Ge}$ and find that a long range chiral order sets in at a transition temperature well below the N{\'e}el temperature ($T_N$). 
Based on the crystalline symmetries, imposed by the chiral magnetic order, we argue for the presence of multiple 
iso-energetic Weyl nodes (nodes that are at same energy and with congruent Fermi surface around them) near chemical potential. 
Using the semi-classical Boltzmann equations, we show that the combined contribution to the net longitudinal magnetoconductance (LMC) and the planar Hall conductance (PHC) from tilted Weyl nodes can lead to signatures, qualitatively distinct from that of a single pair of Weyl nodes.  
In particular, we show that magnetic orders with different chiralities can give rise to different periods in LMC and PHC as a function of the in-plane magnetic field direction. This is ultimately related to differences in the symmetry-imposed constraints on the Weyl nodes.
\end{abstract}

\maketitle

\section{Introduction}\label{sec:intro}
The discovery of large anomalous Hall effect (AHE) at room temperature in Mn$_3$X(Sn,Ge)~\cite{Nakatsuji2015, kiyohara2016giant, Nayak2016AHE} has lead to great interest in these materials from both a fundamental and an application point of view.   
Conventionally, a large AHE is associated with ferromagnetic metals, while $\text{Mn}_3\text{X}$ materials are non-collinear AFMs. 
Thus, the discovery confirms the prediction of AHE in certain non-collinear AFMs~\cite{Chen2014,Kubler2014} due to non-vanishing Berry curvature when certain symmetries are absent. 

Beyond the AHE, other novel transport phenomena, such as large anomalous Nernst effect, spin Nernst effect~\cite{Guo2017} and spin Hall effect~\cite{Zhang2017} have been predicted and subsequently observed in $\text{Mn}_3\text{Sn}$~\cite{Ikhlas2017,Kimata2019} and more recently in Mn$_3$Ge~\cite{Hong2020ANE}. 
Moreover, first-principle calculations predict the existence of Weyl nodes in these materials~\cite{Yang2017Mn3X, Kubler2017}. 
Although these Weyl nodes are not pinned to the chemical potential, the observation of a positive magnetoconductance in the presence of parallel electric and magnetic field provides possible evidence of the chiral anomaly~\cite{Kuroda2017}. 
Thus, these materials are possibly Weyl metals~\cite{Burkov2018}, where the Weyl nodes, though not pinned to the chemical potential, are close enough to it to produce observable transport signatures. 

The use of AFM materials in spintronic devices has several advantages over their ferromagnetic counterparts~\cite{Jungwirth2016,Gomonay2014}. 
To list a few: (i) due to the lack of macroscopic magnetization the information stored in AFM devices is robust against stray fields and magnetic moments of neighboring elements. 
This allows for more compact packing of AFM spintronic devices, facilitating miniaturization. 
(ii) The characteristic frequencies of switching between different AFM states are several orders of magnitude higher than those of FM materials, which leads to faster dynamics~\cite{Fiebig2008}. 
In addition to the inherent advantages, related to AFM character, the non-trivial topology can further enrich the potential applications due to dissipationless transport, associated with the topologically protected boundary states~\cite{Smejkal2018}.
Thus, the interlinked combination of non-collinear antiferromagnetism, non-trivial topology, and novel transport phenomena make these materials of particular interest for spintronic devices. 
It is clear that underlying all these phenomena is the non-collinear magnetic order. 
This magnetic order and its relation to magnetotransport is the focus of this article. 

$\text{Mn}_3\text{X}(\text{Sn},\,\text{Ge})$ has a hexagonal crystal structure (space group $P6_3/mmc$)~\cite{Kren1975}, where 
the magnetic $\text{Mn}^{3+}$ ions arrange in two Kagome layers related by inversion symmetry and separated by half a $c$-axis lattice constant. 
The magnetic $\text{Mn}^{3+}$ ions in the Kagome layer give rise to a $120^{\circ}$ antiferromagnetic order that sets in at a N\'eel temperature $T_N\approx380$~K. 
However, symmetry analysis predicts various candidate $120^{\circ}$ AFM spin configurations, as shown in Fig.~\ref{fig:Mn3Ge_mag_cell} (b)-(e), and unambiguously identifying a unique ground state can be challenging. 
This ambiguity may be partially resolved using spherical neutron polarimetry measurements, which demonstrate that two $E_{1g}$ configurations, shown in Fig.~\ref{fig:Mn3Ge_mag_cell} (d), (e), are equally likely ground states for $\text{Mn}_3\text{Sn}$~\cite{Brown1990}, while $E_{1g}(A_y)$ configuration is the unambiguous ground state for $\text{Mn}_3\text{Ge}$~\cite{Soh2020}. 

Noticeably, the two $E_{1g}$ configurations have the same sense of real-space spin rotation over a triangular plaquette  (chirality)~\footnote{This chirality should be distinguished from the scalar spin chirality $\vec{S}_i\cdot \vec{S}_j\times \vec{S}_k$} and are related by rigid rotation of all the spins. In contrast, the $B_{1g}$ and $B_{2g}$ configurations have real-space spin chirality, opposite to that of the $E_{1g}$ configurations. 
It is known that in presence of spin-orbit coupling, such co-planar configuration can lead to accumulation of Berry phase as an electron moves around the plaquette and can have strong influence on transport~\cite{Zhang2020}. 
Hence for the opposite chirality, one can expect different transport signatures.
Even if a unique zero-field ground state is unambiguously determined, the determination of temperature and magnetic field dependent magnetization is crucial to fully understand possible topological responses such as the negative longitudinal magnetoresistance, photogalvanic effect, and the anomalous Nernst effect. 

Here we present an extensive study of the temperature-dependent magnetic structure of $\text{Mn}_3\text{Ge}$ using finite-temperature classical Monte Carlo and time-dependent integration of Landau-Lifshitz-Gilbert (LLG) equations. Our goal is to establish the temperature-dependent magnetic order and how it affects the electronic structure and therefore the magnetotransport. 
Consistent with the experimental N\'eel temperature, we find that magnetic order starts to set in below $T_N\sim 380 K$ with large fluctuations (Goldstone modes) as the uniaxial anisotropy is very small. 
At higher temperature (but still below $T_N$), these fluctuations suppress long-range real-space chiral order. 
The degeneracy between the two chiralities is lifted due to a finite Dzyaloshinskii-Moriya interactions (DMI) and at lower temperature (around $200$~K and below) long-range chiral order is established,  consistent with the $E_{1g}$ configuration. 
Using the symmetries, imposed by these magnetic configurations, incorporated in a model Weyl metal, we discuss longitudinal magneto conductivity (LMC) and planar Hall conductivity (PHC) and show how a chirality-dependent switch in their periodicity might occur due to symmetry-constrained motion of the Weyl nodes. 
In particular, we show that while $B_{1g}$ and $B_{2g}$ configurations exhibit two-fold periodicity, the $E_{1g}$ configuration can lead to higher-order periods.
Interestingly, a recent temperature-dependent measurement of the longitudinal magnetoresistance with the current and magnetic field in the crystallographic $ab$ plane and the magnetic field at some angle $\varphi$ to the current in high-quality Mn$_3$Ge thin films reveals a striking temperature dependence of the magnetoresistance on the field angle~\cite{hong2021epitaxial,Priv_Comm_Anand}, which is consistent with our main theoretical findings. 

The paper is organized as follows. In Sec.~\ref{sec:methods}, we describe the magnetic model and the computational methods and details. In Sec.~\ref{sec:results}, we discuss our main numerical results of temperature dependent magnetism in detail. In Sec.~\ref{Sec:AMR-PHE}, we discuss the calculation of LMC and PHC, using a model of a Weyl metal, which has the symmetry, imposed by the magnetic orders, discussed in Sec.~\ref{sec:results}. We end with a summary of the results and concluding remarks in Sec.~\ref{Sec:Disc}.

\section{Model and Methods}\label{sec:methods}
\begin{figure}
    \centering
    \includegraphics[width=0.5\textwidth]{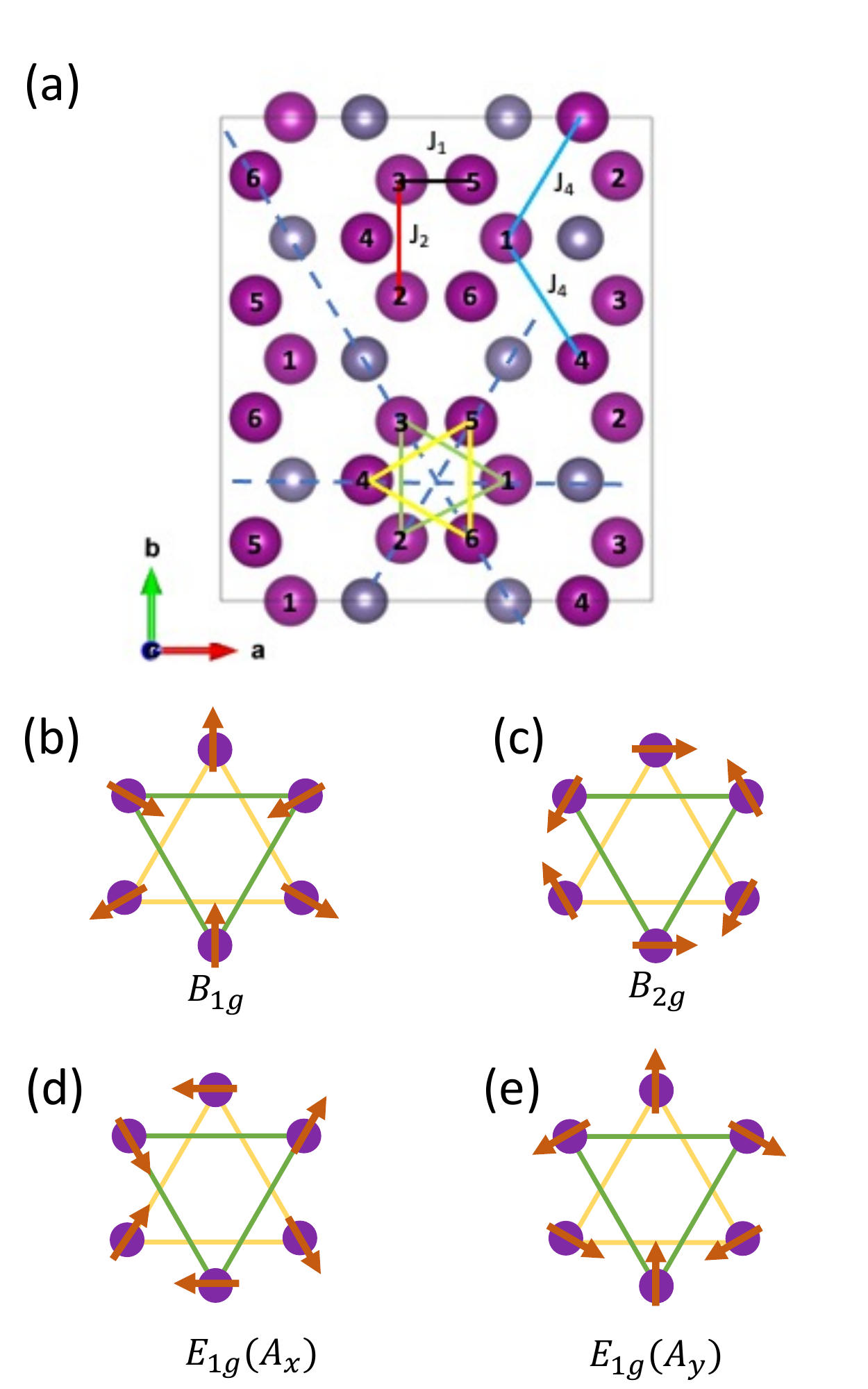}
    \caption{(a) Orthorombic Mn$_3$Ge unit cell used in the magnetic simulations (Crystal structure produced using VESTA~\cite{Momma2011}). The unit cell contains 24 Mn atoms on six different sublattices labeled 1 - 6. Sublattices 1 - 3 is on one crystallographic $c$-plane, and sublattices 4 - 6 on a different plane; the two sublattices form hexagons projected on the $ab$ plane as indicated in the lower part of the figure with DMI coupling atoms 1-2-3 and atoms 4-5-6, as indicated by yellow and green triangles. In-plane AFM near-neighbor coupling $J_2$, inter-plane near-neighbor AFM and inter-plane FM couplings $J_1$ and $J_4$ are indicated with red, black, and blue lines, respectively. Single-site uniaxial anisotropy directions are indicated by the dashed blue lines. (After Chen {\em et al.}\cite{chen2020antichiral}.) 
    (b)-(e) Different localized magnetic moment configuration allowed by $120^{\circ}$ AFM order.
    }
    \label{fig:Mn3Ge_mag_cell}
\end{figure}

For the atomistic simulations of magnetic order as function of temperature, we start with the model Heisenberg Hamiltonian $H$ described by Chen {\em et al.}~\cite{chen2020antichiral} (see Fig.~\ref{fig:Mn3Ge_mag_cell}). The model describes local moments located on the Mn sites with intra-plane near-neighbor AFM couplings $J_2$ between near-neighbor Mn sites on the same crystallographic $c$ plane, inter-plane AFM couplings $J_1$ and inter-plane FM couplings $J_4$; in addition, there is an intra-plane near-neighbor DMI coupling $D$ with the unit DMI vector ${\mathbf d}_{ij}$ connecting spins $i$ and $j$  (indicated by yellow and green triangles in Fig.~\ref{fig:Mn3Ge_mag_cell}) along the $c$ axis, and a weak single-site uniaxial anisotropy with easy axes indicated by the dashed lines in Fig.~\ref{fig:Mn3Ge_mag_cell}. We use a Cartesian coordinate system with the $x$- and $y$-axes along the crystallographic $a$ and $b$ axes, and the $z$ axis along the crystallographic $c$ axis (out of the plane in Fig.~\ref{fig:Mn3Ge_mag_cell}, and write
\begin{align}
    & H_M  =  \frac{J_1}{2}\sum_{1,ij}{\mathbf S}_i\cdot{\mathbf S}_j
    +\frac{J_2}{2}\sum_{2,ij}{\mathbf S}_i\cdot{\mathbf S}_j
    +\frac{J_4}{2}\sum_{4,ij}{\mathbf S}_i\cdot{\mathbf S}_j \notag\\
    &+\frac{D}{2}\sum_{D,ij}{\mathbf d}_{ij}\cdot\left ({\mathbf S}_i\times{\mathbf S}_j\right )
   -k\sum_i\left ({\mathbf S}_i\cdot\hat n_i\right )^2 \notag\\
   &-2.5\mu_B\sum_i{\mathbf S}_i\cdot{\mathbf B}_{\rm ext},\
    \label{eqn:heisenberg}
\end{align}
where ${\mathbf S}_i$ is the spin director at site $i$, the sums $1,ij$ {\em etc} indicates a sum over sites connected by the coupling $J_1$, $n_i$ is the anisotropy easy axis direction for site $i$, and ${\mathbf B}_{\rm ext}$ is an applied uniform external induction field.

We used an orthorombic unit cell with 24 Mn atoms shown in Fig.~\ref{fig:Mn3Ge_mag_cell} and the Vampire code~\cite{evans2014atomistic} with typically a $4.4863$~nm$\times5.1803$~nm$\times2.0942$~nm supercell containing 3,000 atoms with periodic boundary conditions in most simulations, although we also used larger supercells up to 8.9726$\times$10.36066$\times$4.18841~nm$^3$ to ensure that our results were not affected by the choice of supercell. We employed both finite-temperature Monte Carlo (MC) simulations as well as direct integration of the stochastic Landau-Lifshitz-Gilbert (s-LLG) equation\cite{garcia1998langevin}  both with and without atomistic dipolar interactions in the s-LLG, for the magnetization dynamics at finite temperatures, typically monitoring sublattice magnetization at the six sublattices denoted in Fig.~\ref{fig:Mn3Ge_mag_cell}. For thermal averages we would collect between 50,000 and 1,000,000 Monte Carlo steps per spin (MC simulations) or timesteps at 0.1~fs (s-LLG simulations). We used a dimensionless damping $\alpha$ in the s-LLG of $\alpha=0.1$. This is typically a good choice for modeling equilibrium properties; we also note that this is not inconsistent with an observed large magnon damping in Mn$_3$Ge\cite{Cable1993,chen2020antichiral}. We could not detect any discernible differences in the results obtained using MC or s-LLG with our without dipolar interactions. Instead, the main issue in the simulations was always to ensure that the system did not get trapped in a local equilibrium, and we carefully montitored instantaneous average energies at each temperature and field as well as sublattice magnetizations to eliminate simulations that were clearly trapped in a local equilibrium. Starting with the coupling constants given by Chen {\em et al.}~\cite{chen2020antichiral}, we first verified the N\'eel temperature at 365~K using standard Monte Carlo simulations with averages over 50,000 Monte Carlo steps per spin (MCS), slightly adjusting the coupling constants by about 10\% from the values given by Chen {\em et al.}~\cite{chen2020antichiral} (Table~\ref{tab:constants}). Note that in Eq.~(\ref{eqn:heisenberg}) we double-count the sites in the summations and correct with a factor of $1/2$ which leads to a factor of 2 different from the values quoted by Chen {\em et al.}~\cite{chen2020antichiral}. As is seen in Table~\ref{tab:constants}, $J_2$ and $J_4$ are much larger and control the N\'eel temperature; the anisotropy constant corresponds to a temperature of about 0.01~K and plays no role at all in the magnetic order down to temperatures $\sim0.1$~K. This implies that the energy is continuously degenerate under in-plane rotations of all spins as the Hamiltonian only depends on relative angles of spins when the anisotropy can be ignored. The DMI constant $D$ lifts the degeneracy between chiralities in a basic plaquette defined by sites 1 - 3 in Fig.~\ref{fig:Mn3Ge_mag_cell}. We shall see that even though $D$ is small it affects the system on large temperature scales, $\sim100$~K, as the coupling $J_4$ connects chiralities in different plaquettes. 

We integrated the s-LLG equation for typically $10^7$ time steps in steps of 0.1~fs at fixed temperatures from 0.1~K to 400~K, first randomizing the spins for $10^6$ time steps at $T=300$~K, after which we sampled sub-lattice spins every ten time steps. The simulations were run with no external field and with a 14~T field along the $a$-axis (Fig.~\ref{fig:Mn3Ge_mag_cell}) and at $30^\circ$, $45^\circ$, $60^\circ$, $75^\circ$, and $90^\circ$ relative to the $a$-axis. 

\begin{table}
 \caption{Coupling parameters for the magnetic model after Ref.~[\onlinecite{chen2020antichiral}]. The couplings are identified in Fig.~\ref{fig:Mn3Ge_mag_cell} (a).}
\begin{center}\label{tab:constants}
 \begin{tabular}{|c | c | c | c | c |} 
 \hline
$J_1$~(meV) & $J_2$~(meV) & $J_4$~(meV) & $k$~(meV)& $D$~(meV)\\ [0.5ex] 
 \hline\hline
 $6.24\times10^{-3}$ & 37.4 & -18.7 & $1.0\times10^{-3}$ &$2.18\times10^{-2}$ \\ 
 \hline
\end{tabular}
\end{center}
\end{table}

\section{Results and Discussion}
\label{sec:results}
\subsection{Magnetic modeling: finite-temperature Monte Carlo and stochastic Landau-Lifshitz-Gilbert simulations}\label{subsec:magnetic}
\begin{figure*}
    \centering
    \includegraphics[width=1.7\columnwidth]{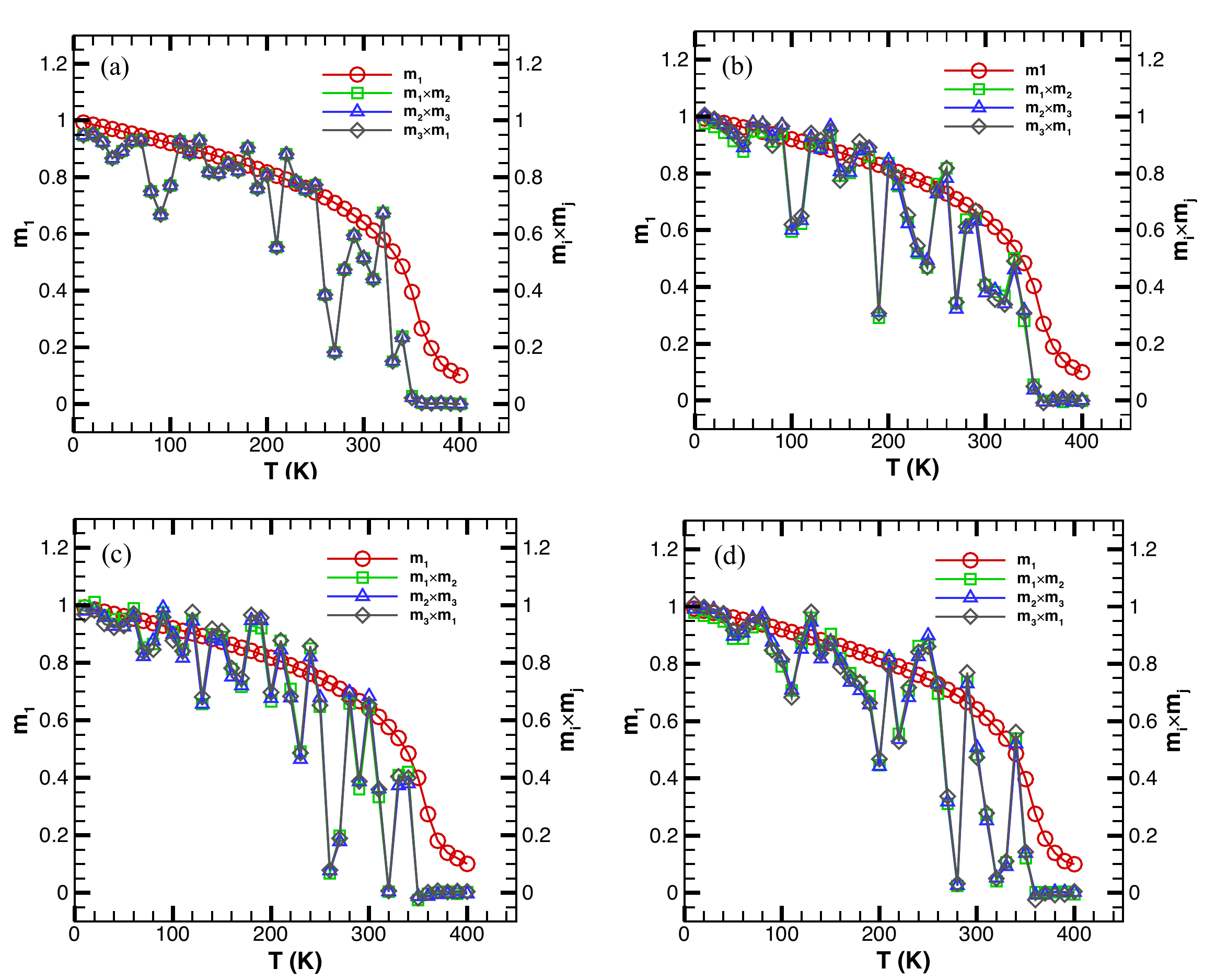}
    \caption{Thermal average of the magnetization magnitude on sublattice 1 (red line and circle), and normalized mean-field chirality as a function of temperature for (a) no magnetic field, and 14~T field along (b) the $a$ axis, (c) $30^\circ$ to the $a$ axis, and (d) $60^\circ$ to the $a$ axis. The different colors (green line and square, blue line and triangle, and grey line and diamond) denote the three different ways the chirality $\hat z\cdot({\mathbf S}_i\times{\mathbf S}_j)$ can be calculated in an elementary triangular plaquette.}
    \label{fig:chirality}
\end{figure*}

As the temperature is decreased below the N\'eel temperature, order sets in on the sub-lattices with the spins on a basic plaquette of three spins arranged in the classic in-plane N\'eel AFM order with an in-plane rotation of $120^\circ$ between consecutive spins and the spin on a basic hexagon (sublattices 1 - 6 in Fig.~\ref{fig:Mn3Ge_mag_cell}) arranged in an octupolar order that transforms as the irreducible representation $B_{2g}$ or $E_{1g}$ (Fig.~\ref{fig:Mn3Ge_mag_cell} (b) - (e)). The main difference between the two representations is that they have different chiralities: the $E_{1g}$ order has positive chirality (as we move in the counter-clockwise direction around a basic plaquette of three spins, the spins rotate counter-clockwise) and the $B_{2g}$ order has negative chirality. We calculate the magnetic chirality as $\hat z\cdot[\vec{S}_i\times\vec{S}_j]$, where $\vec{S}_i$ and $\vec{S}_j$ are two spins in an elementary triangular plaquette, with the order of $i$ and $j$ going counterclockwise. Because of the DMI, these two chiral orders are not degenerate, but at $T\agt250$~K thermal fluctuations destroy the long-range chiral order. As the temperature is lowered to below approximately 250~K, specific chiral order develops. Depending on initial conditions, any simulation at temperatures below 200~K or so would develop a definite chirality, but the chirality would differ from simulation to simulation. Figure~\ref{fig:chirality} depicts the  normalized mean-field chirality calculated as $2\hat z\cdot\mathbf{m}_i\times\mathbf{m}_j/\sqrt{3}$, with $i,j=1,2$, $i,j=2,3$, and $i,j=3,1$, and $\mathbf{m}_i$ the thermal averages of the sublattice magnetization obtained in MC simulations with 100,000 MC steps per spin. The figure also shows the thermal average of the magnetization magnitude on sublattice 1 (it is the same on the other sublattices). The figure shows that sublattice magnetization magnitude goes to zero at about $T_N\approx365$~K, with the typical finite-size rounding near $T_N$. In contrast, the mean-field chirality drops sharply to zero at temperatures well below $T_N$, in a range of temperatures from 280~K to 340~K (in the simulations depicted in Fig.~\ref{fig:chirality}, this range is about 340~K to 360~K). Furthermore, the mean-field chirality exhibits large fluctuations and varies considerably as function of temperature for $T\agt200$~K. We have also observed the chirality to change sign at temperatures $T\agt200$~K. This leads us to conclude that a robust chiral order sets in at temperatures $T\alt200$~K, well below $T_N$, but that the chiral order is not a distinct order but a feature of the N\'eel order in this system. We estimate the energy difference between the two chiralities to be 0.08~meV/spin. This a little larger than the DMI coupling of about 0.02~meV, but substantially smaller than the main couplings $J_1$ and $J_2$.  However, the fact that low-temperature simulations would be trapped in a specific chirality at temperatures much above $D/k_B$ clearly indicates that the barrier for switching between chiralities is substantially larger than the energy difference, about $200$~K, corresponding to about 20~meV. This is close to the ferromagnetic coupling $J_4$: when long-range chiral order has developed, the energy cost to switch chirality involves breaking the $J_4$ coupling, which is responsible for stabilizing long-range chiral order.

\begin{figure*}
    \centering
    \includegraphics[width=1.0\textwidth]{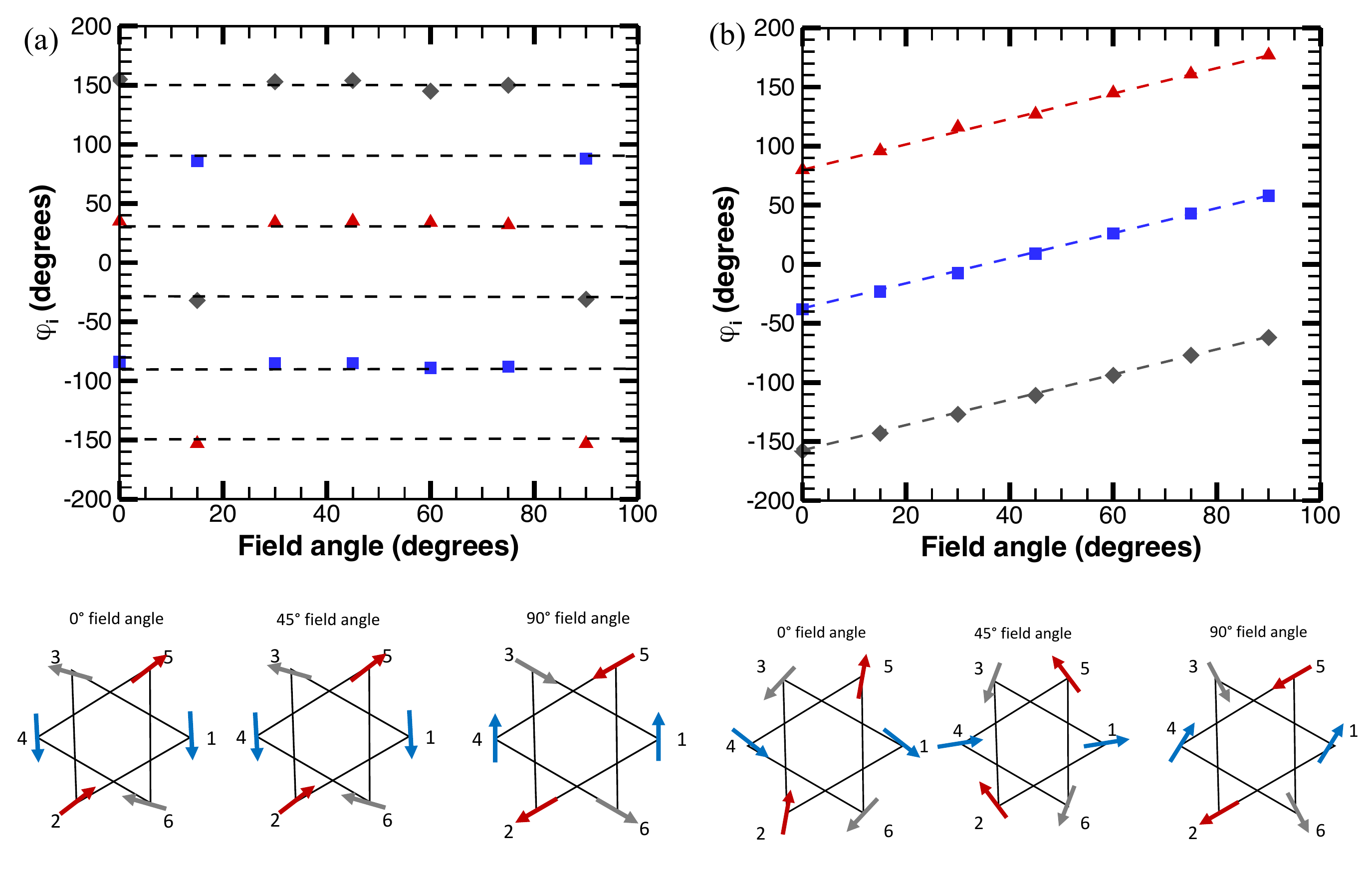}
    \caption{Top row: Sublattice magnetization as function of applied magnetic field angle at $B=14$~T and $T=10$~K for (a) no magnetic defects, and (b) a small substitutional Mn defect on a Ge site, with sublattice 1 in blue, 2 in red, and 3 in gray; the orange dots and dashed line indicate a direction parallel to that of the applied field. The bottom row illustrates corresponding sublattice spin configurations at three different field angles. Without a magnetic defect, the low-temperature (ground state) spin configuration is the positive-chirality $E_{1g}$ spin configuration (Fig.~\ref{fig:Mn3Ge_mag_cell}). 
    With a magnetic defect present, the spins rotate uniformly with the magnetic field direction.}
    \label{fig:mag_angle}
\end{figure*}

We also examined in some detail the response of the magnetization to an external applied field. It was suggested by Kiyohara \textit{et al.}~\cite{kiyohara2016giant} that the spin configuration in a basic hexagon (sublattice sites 1 - 6) will rotate easily when an applied external field is rotated in-plane due to the very small anisotropy. On the other hand, an observation made by Chen {\em et al.}~\cite{chen2020antichiral} was that in the presence of an external magnetic field, a small {\em transverse} ferromagnetic component of about $0.2$~$\mu_B$ per Mn$_3$Ge unit is developed in response to the applied field. In our simulations, we do not see any evidence of spin rotation with the applied field direction. In fact, even at an applied field of 14~T, the sublattice magnetization direction is largely independent of the applied magnetic field. There is in general a small component of magnetization, less than 0.01$\mu_B$ per Mn at $T=10$~K, induced by the magnetic field, that is along the direction of the magnetic field, with a deviation from the direction of the applied field of $1^\circ$ or less. Figure~\ref{fig:mag_angle} depicts the angle of the sublattice magnetization relative to the $a$ axis for an external field of 14~T applied in plane at different angles to the $a$ axis at $T=10$~K obtained simulating the s-LLG and averaging over 1,000,000 time steps at 0.1~fs. For systems without any defects, neither MC nor the s-LLG simulations show any evidence of rotation of the sublattice magnetization. This is particularly clear in the s-LLG simulations; the MC simulations are, as is typically the case for these systems, less unequivocal. It is possible that extrinsic factors are responsible for observed or inferred rotations of the magnetization with external field, in particular magnetic defects. In fact, Mn$_3$Ge is stable only when there is an excess of Mn, typically about 10\%, occupying Ge sites\cite{kiyohara2016giant,chen2020antichiral}. Such defects can couple ferro- or antiferro-magnetically to nearby Mn ions, and there may also be longer-range Ruderman–Kittel–Kasuya–Yosida interactions that can be either ferromagnetic or antiferromagnetic. We chose a simple model in order to explore the effect of Mn substitutions on Ge sites. We substituted one Ge atom with a Mn atom and coupled it ferromagnetically to its nearest Mn neighbors. This corresponds to a defect concentration of about 4\%. In this case, the simulations show a clear rotation of the sublattice magnetizations in response to rotation of the external field. In addition, there is now consistently a larger deviation, about $2^\circ$ to $3^\circ$, between the induced moment and the direction of the applied field. The direction of the deviation also depends on the chirality of the system. While the magnitude of the induced transverse magnetic moment is much smaller than the transverse magnetization observed by Chen {\em et al.}\cite{chen2020antichiral},  this suggests that the observed transverse magnetization arises from interaction with defects. Therefore, we believe that observed spin rotation with magnetic field as well as transverse components are likely the result of Mn defects on Ge sites. We note that the transverse moment we observe is much smaller than that observed by Chen {\em et al.}, but the defect concentration in our simulations is also smaller. It is not inconceivable that the effect of increasing defect density is not simply additive.
Moreover, in our calculations the spins rotate along the magnetic field rotation directions. 
This is in contrast to Ref.~\cite{kiyohara2016giant} and Ref.~\cite{chen2020antichiral}, where spin rotation is in direction opposite to the  magnetic field rotation. 
We remark that in our calculations, the extra $\text{Mn}$ atom couples ferromagnetically with its nearest neighbors. 
It is possible that for antiferromagnetic coupling, or with the inclusion of longer-range interactions, the spin rotation is opposite sense to that of the magnetic field.

\subsection{Magnetotransport\label{Sec:AMR-PHE}}
Given the results from our finite-temperature simulations of the magnetic order, we now discuss how the configuration of local magnetic moments in $\text{Mn}_3\text{X}$ can influence the magnetotransport. Our focus is on the period of LMC and PHC as functions of the in-plane magnetic field direction. 
The observation of a large AHE~\cite{Nayak2016AHE,kiyohara2016giant} and anomalous Nernst effect~\cite{Hong2020ANE} in Mn$_3$Ge has been attributed to Weyl nodes. The presence of Weyl nodes near the chemical potential has also been shown in density functional theory calculations~\cite{Kuebler2017Mn3X,Yang2017Mn3X}. 
Similar to the conventional anisotropic magnetoresistance (AMR) in a ferromagnetic metal~\cite{mcguire1975anisotropic}, a pair of isotropic Weyl nodes can lead to an anisotropic LMC that has a $\cos^2\varphi$ dependence as a function of field-current angle $\varphi$, when the applied magnetic field, current and the Weyl node pairs all lie in same plane~\cite{Nandy2017PHE}. Unlike conventional AMR, the chiral anomaly may be the reason behind these phenomena in a Weyl metal~\cite{Nandy2017PHE, Burkov2017PHE}. 
If the Weyl nodes are tilted, the anisotropic LMC can develop a further $\cos\varphi$ component that can be the dominant component for the appropriate choice of Weyl node dispersion~\cite{Ma2019PHE}. 
As a result the anisotropic LMC and closely related PHC can show a $2\pi$-period in $\varphi$. 

Here, using the specific symmetries of $\text{Mn}_3X (\text{Ge}/\text{Sn})$, we show that chiral anomaly induced LMC and PHC can also have more complex angular dependence. 
Starting with a simple model of a generic Weyl metal with tilted Weyl cones, we first derive the expression for the LMC and PHC as functions of the magnetic field direction. 
Then, using the symmetries of $\text{Mn}_3\text{X}$ crystal, we enforce the corresponding constraints on the iso-energetic Weyl nodes and the Fermi surfaces around these nodes. Using these symmetry arguments, we show how the LMC and PHC signals and their periods can depend on temperature in these materials. 

\begin{figure*}[t]
  \includegraphics[width=0.8\textwidth]{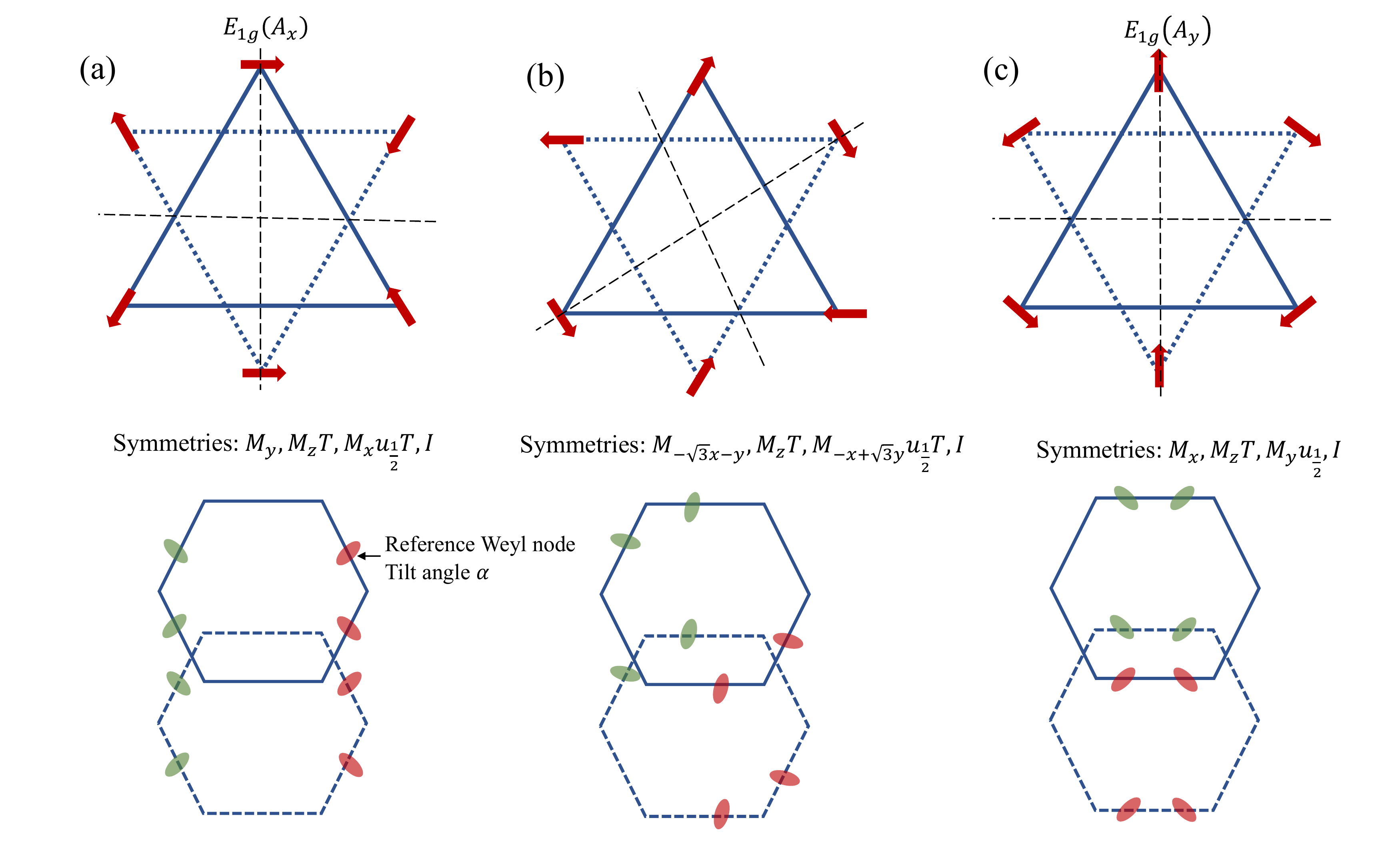}
  \caption{\label{Fig:KagomeSchematic}
  A schematic representation of location of Weyl nodes as a function of localized spins for some high symmetry directions. 
  The bottom panel shows all the symmetry-related Weyl nodes. The red and blue ovals represent Fermi surface around opposite chirality Weyl nodes with tilt. 
  As all the localized spins are rotated simultaneously, the relevant label of crystalline symmetries change, which leads to movement of Weyl points. 
}
\end{figure*}

We start with a simple $\vec{k}\cdot\vec{p}$ Hamiltonian
\begin{align}\label{Eq:Weyl_kp}
    & H_{\chi}(\vec{k}) = \chi\hbar v\vec{k}\cdot\vec{\sigma} + \hbar \vec{t} \cdot\vec{k}\sigma_0\, ,
\end{align}
where $\chi = \pm$ denotes the chirality of the Weyl node, $v$ is the Fermi velocity, $\vec{t}$ is the tilt vector and $\vec{\sigma}$ are the Pauli matrices. We will only consider type-$I$ nodes, which can be ensured by choosing $|\vec{t}| < v$.
Under an external magnetic field $\vec{B}$ and ignoring Landau quantization, the modified band energy 
\begin{align}\label{Eq:bandE}
    & \epsilon^{\chi}_{\tau}(\vec{k}) = \hbar \vec{t}_{\chi} \cdot\vec{k} + \tau \hbar v \vec{k} - g \vec{m}^{\chi}_{\tau,\vec{k}}\cdot\vec{B} \, ,
\end{align} 
that takes the orbital magnetization correction into account, where $g$ is the orbital magnetization $g$-factor and 
\begin{align}\label{Eq:OM}
   \vec{m}^{\chi}_{\tau, \vec{k}} = -\chi e v\frac{\tau \vec{k}}{2|\vec{k}|^2}\, ,
\end{align}
is the orbital magnetic moment. Here $\tau=\pm$ denote the conduction and the valence band respectively. In our analysis, we take $\vec{B} = B(\cos\varphi , \, \sin\varphi ,\, 0)$ and $\vec{t} = t (\cos\vartheta , \, \sin\vartheta , \, 0)$ in the $xy$ plane (the $ab$-plane of $\text{Mn
}$ kagome layers). 

Keeping the current direction fixed along the $x$-axis, for a pair of opposite chirality Weyl nodes lying in the $xy$-plane, the final expressions for the conductivity can be summarized as follows (derivation is presented in the appendix~\ref{App:LMC_PHC_derivation}):
\begin{subequations}\label{Eq:cond} 
\begin{align}
    \sigma^{\chi}_{xx} = \sum^{2}_{n=0} \sum^{4}_{m = -4 } a^{\chi}_{n,m} \cos (n\varphi +  m\vartheta),\, \\
    \sigma^{\chi}_{xy} = \sum^{2}_{n=0} \sum^{4}_{m = -4 } b^{\chi}_{n,m} \sin (n\varphi +  m\vartheta) ,
\end{align}
\end{subequations}
where $n,m$ take integer values. 
The coefficients $a^{\chi}_{n,m}$ and $b^{\chi}_{n,m}$ depend on the magnetic field strength and on the band parameters such as $\mu, \vec{t}, v$, and are listed in Eq.~\ref{Eq:App_coeff_xx} and Eq.~\ref{Eq:App_coeff_xy}. In deriving Eq.~\ref{Eq:cond}, we have extended the results of Ref.~\cite{Ma2019PHE} for arbitrary tilt direction and included the orbital magnetization corrections explicitly. 
We now incorporate the symmetries of the different magnetically-ordered states in $\text{Mn}_3\text{X}$ in our transport calculations. 

The top part of Fig.~\ref{Fig:KagomeSchematic} (a), shows the $E_{1g}(A_x)$, magnetic configuration, where the vertices of the triangles represent the magnetic $\text{Mn}$-ions. This particular structure has inversion symmetry ($I$) along with mirror symmetries $M_y$, $M_z T$, and $M_x u_{1/2} T$, where $T$ is the time reversal and $u_{1/2}$ is the translation by half lattice constant in $z$-direction. 
Next, we assume that the Fermi level lies near a reference Weyl node as shown in bottom part of  Fig.~\ref{Fig:KagomeSchematic} (a), with chirality $\chi = +$ and tilt $\vec{t} = t(\cos\vartheta,\, \sin \vartheta ) $. 
Based on the symmetries listed above, all iso-energetic Weyl nodes are shown in the bottom part of Fig.~\ref{Fig:KagomeSchematic} (a) with the red and green ellipses representing opposite chirality and the major axis of the ellipse along the tilt direction. 
By using a small $\vec{k}$-expansion around each of these Weyl nodes, we can write down a simple $\vec{k}\cdot\vec{p}$ Hamiltonian of the type shown in Eq.~(\ref{Eq:Weyl_kp}), with the details listed in Table~\ref{Table:2}.
\begin{widetext}
\begin{table*}
\caption{Effective $\vec{k}\cdot\vec{p}$ near each Weyl node}
\begin{center}\label{Table:2}
 \begin{tabular}{|c | c | c | c | c | c|} 
 \hline
 Symmetry & $\vec{k}= (k_x,k_y,k_x)$ & $\chi$ & $\vec{\Omega}=(\Omega_x,\Omega_y,\Omega_z)$  & $\vec{t} = (t_x,t_y)$ & $H = \chi\hbar v \vec{k}\cdot\vec{\sigma} + \hbar (t_xk_x+ t_yk_y)$\\ [0.5ex] 
 \hline\hline
 $M_y$ & $(-k_x,k_y,k_z)$ & $-\chi$ & $(\Omega_x,-\Omega_y,-\Omega_z)$ & $(-t_x,t_y)$  & $-\chi\hbar v \vec{k}\cdot\vec{\sigma} + \hbar (-t_xk_x+ t_yk_y)$\\ 
 \hline
 $M_zT$ & $(-k_x,-k_y,k_z)$ & $-\chi$ & $(\Omega_x,\Omega_y,-\Omega_z)$ & $(-t_x,-t_y)$ & $-\chi\hbar v \vec{k}\cdot\vec{\sigma} + \hbar (-t_xk_x - t_yk_y)$\\
 \hline
 $M_xu_{1/2}T$ & $(-k_x,k_y,-k_z)$ & $-\chi$ & $(\Omega_x,-\Omega_y,\Omega_z)$  & $(-t_x,t_y)$  & $-\chi\hbar v \vec{k}\cdot\vec{\sigma} + \hbar (-t_xk_x+ t_yk_y)$\\
 \hline
 $I$ & $(-k_x,-k_y,-k_z)$ & $-\chi$ & $(\Omega_x,\Omega_y,\Omega_z)$ & $ (-t_x,-t_y)$ & $-\chi\hbar v \vec{k}\cdot\vec{\sigma} + \hbar (-t_xk_x - t_yk_y)$ \\
 \hline
\end{tabular}
\end{center}
\end{table*}
\end{widetext}

The magnetic configuration in the top part of Fig.~\ref{Fig:KagomeSchematic} (b) can be obtained by starting from $E_{1g}(A_x)$ configuration and applying a counter-clockwise rotation of all the localized spins by  $\pi/3$. As a result the mirror planes perpendicular the triangle rotate by the same angle but clockwise.  
The remaining symmetries, \textit{i. e.} $I$ and mirror in the triangular plane ($M_zT$) symmetries are still intact. 
The configuration in Fig.~\ref{Fig:KagomeSchematic} (b) can also be viewed as an overall counter-clockwise rotation of the crystal by $2\pi/3$ followed by time reversal. 
As a result the Weyl nodes move to different positions in the momentum space as shown in the bottom panel of Fig.~\ref{Fig:KagomeSchematic} (b). Similarly, Fig.~\ref{Fig:KagomeSchematic} (c) shows when all spins are rotated by $\pi/2$ counter-clockwise, the perpendicular mirror planes rotate by $\pi/2$-clockwise. 

\begin{figure}[!htb]
  \includegraphics[width=0.5\textwidth]{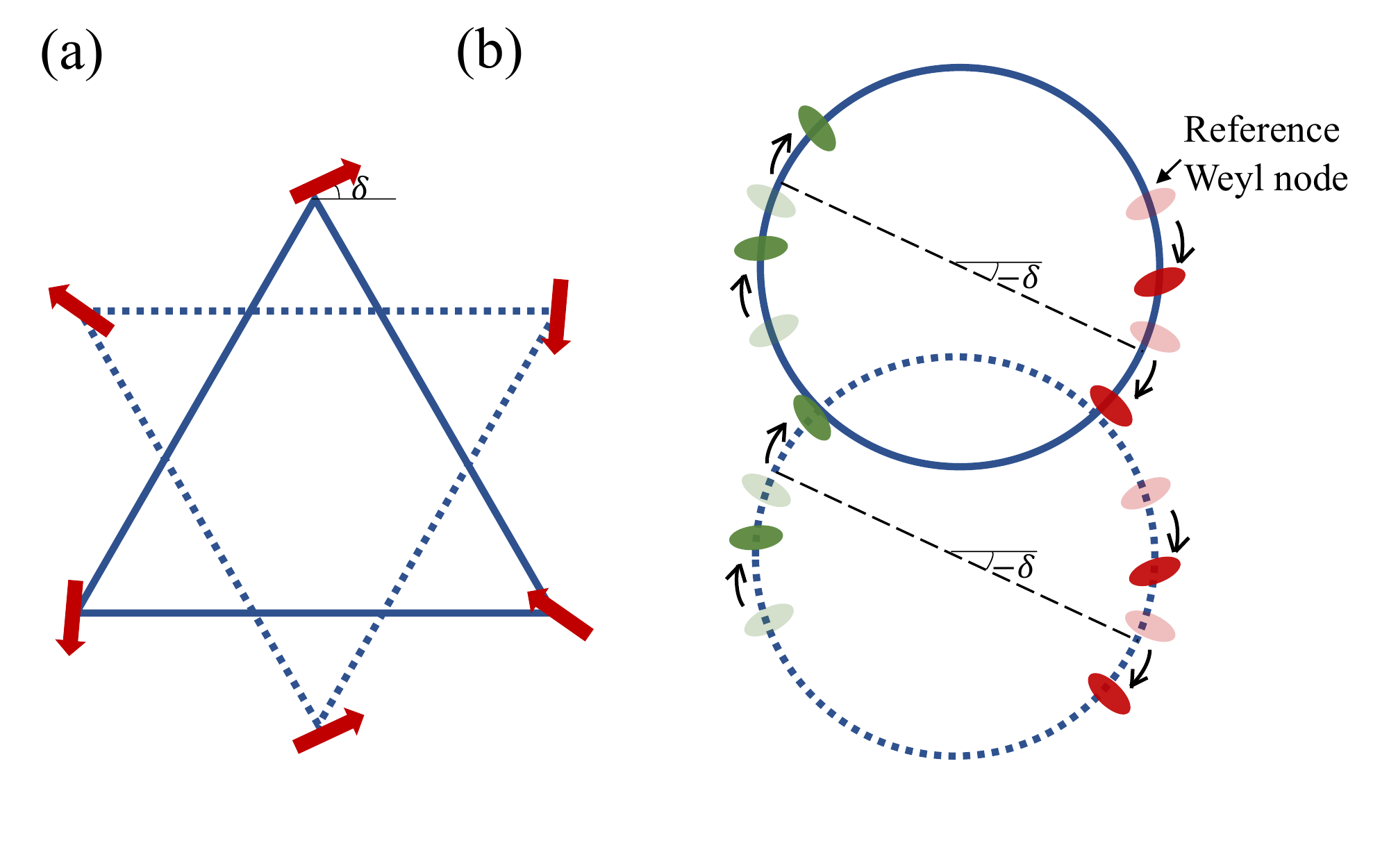}
  \caption{\label{Fig:Rotation}
  A schematic representation of location of Weyl nodes as a function of localized spins, (a) Rotation of localized spins, (b) Corresponding movement of Weyl nodes. The shaded ovals are at the original location of Weyl nodes and the solid ovals are at the new locations. As all the localized spins are rotated simultaneously, the Weyl nodes move because of change in mirror planes.  
}
\end{figure}

From the above analysis, it is clear that as the localized spins are rotated in a rigid fashion, the Weyl nodes must move in \vec{k}-space. Figure~\ref{Fig:KagomeSchematic} only shows configurations in which spins are aligned along some high symmetry directions. However, if the spins are rigidly rotated in an arbitrary direction, the above-mentioned symmetries, apart from the $I$-symmetry, may all be broken. Hence, in principle, the movement of Weyl nodes as function of spin rotation can be complicated. In the analysis below, we consider the simplest possible picture of Weyl nodes movement shown schematically in Fig.~\ref{Fig:Rotation}. 
We note that the positions of the Weyl nodes when the spins are aligned along high-symmetry directions are determined by symmetry irrespective of the particular model.
As every localized spin is rotated by an angle $\delta$, the movement of Weyl nodes can be thought of by rotating the momentum space around the origin by $-\delta$. The reference Weyl node as pointed out in Fig.~\ref{Fig:KagomeSchematic} (a) is now at a new tilt angle $\vartheta-\delta $. Given this, next, we add the contribution to the conductivity from all the eight iso-energy Weyl nodes according to Eq.~\ref{Eq:cond}, and after some manipulation obtain
\begin{subequations}\label{Eq:Conductivity_total}
\begin{align}
    & \sigma_{xx} = \sum^{6}_{n=0} \bar{a}_n \cos (n\varphi) \\
    & \sigma_{xy} = \sum^{6}_{n=1} \bar{b}_n \sin (n\varphi) .
\end{align}
\end{subequations}
Here, we have assumed that the spin rotation $\delta$ either follows the magnetic field rotation $\varphi$ ( i.e. $\delta = \varphi $) as we find in our case or it is opposite to magnetic field rotation (i.e. $\delta = -\varphi$) as found in Ref.~\cite{kiyohara2016giant} and Ref.~\cite{chen2020antichiral}. 
Depending on the situation, we obtain the coefficients
\begin{widetext}
\begin{subequations}\label{Eq:Cond_tota_coeff}
\begin{align}
    & \bar{a}_n = \begin{cases}
    4\sum'_m \sum_{\chi=\pm} \chi^{m} (a^{\chi}_{n+m,m} + a^{\chi}_{m-n,m})\cos (m \vartheta ) & \text{$\delta = \varphi$}\\
    4\sum'_m \sum_{\chi=\pm} \chi^{m} (a^{\chi}_{n-m,-m} + a^{\chi}_{-m-n,-m})\cos (m \vartheta ) & \text{$\delta = -\varphi$}\\
    \end{cases}\\
    & \bar{b}_n = \begin{cases}
    4\sum'_m \sum_{\chi=\pm} \chi^{m} (b^{\chi}_{n+m,m} - b^{\chi}_{m-n,m})\cos (m \vartheta ) & \text{$\delta = \varphi$}\\
    4\sum'_m \sum_{\chi=\pm} \chi^{m} (b^{\chi}_{n-m,-m} - b^{\chi}_{-m-n,-m})\cos (m \vartheta ) & \text{$\delta = -\varphi$}.
    \end{cases}
\end{align}
\end{subequations}
\end{widetext}

Here the $\sum'$ sums over the certain values of $m \in [-4,\, 4]$, such that the first subscript $p \in [0,2 ]$ for  $a^{\chi}_{p,m}$ and $b^{\chi}_{p,m}$. 
The most important observation is that the higher order (up to six-fold) components generically emerge in the angular dependence of the LMC and PHC signals as a function of the magnetic field direction $\varphi$. 
This dependence is a direct consequence of the movement of tilted Weyl nodes in $\vec{k}$-space. 
The movement of Weyl nodes in $\vec{k}$-space is ensured because of change in the underlying symmetries due to spin rotation. 
Such a spin rotation in response to an external field has been suggested as a consequence of coupling of the octupolar order to the external field~\cite{chen2020antichiral,Kimata2021}. We did not observe that in our finite-T simulations without disorder. We note however that our simple model of substitutional disorder of Mn on Ge sites [see Fig.~\ref{fig:mag_angle} (b)] suggests that substitutions of Mn on Ge sites can lead to such rotations .

\begin{figure}[h]
  \includegraphics[width=0.45\textwidth]{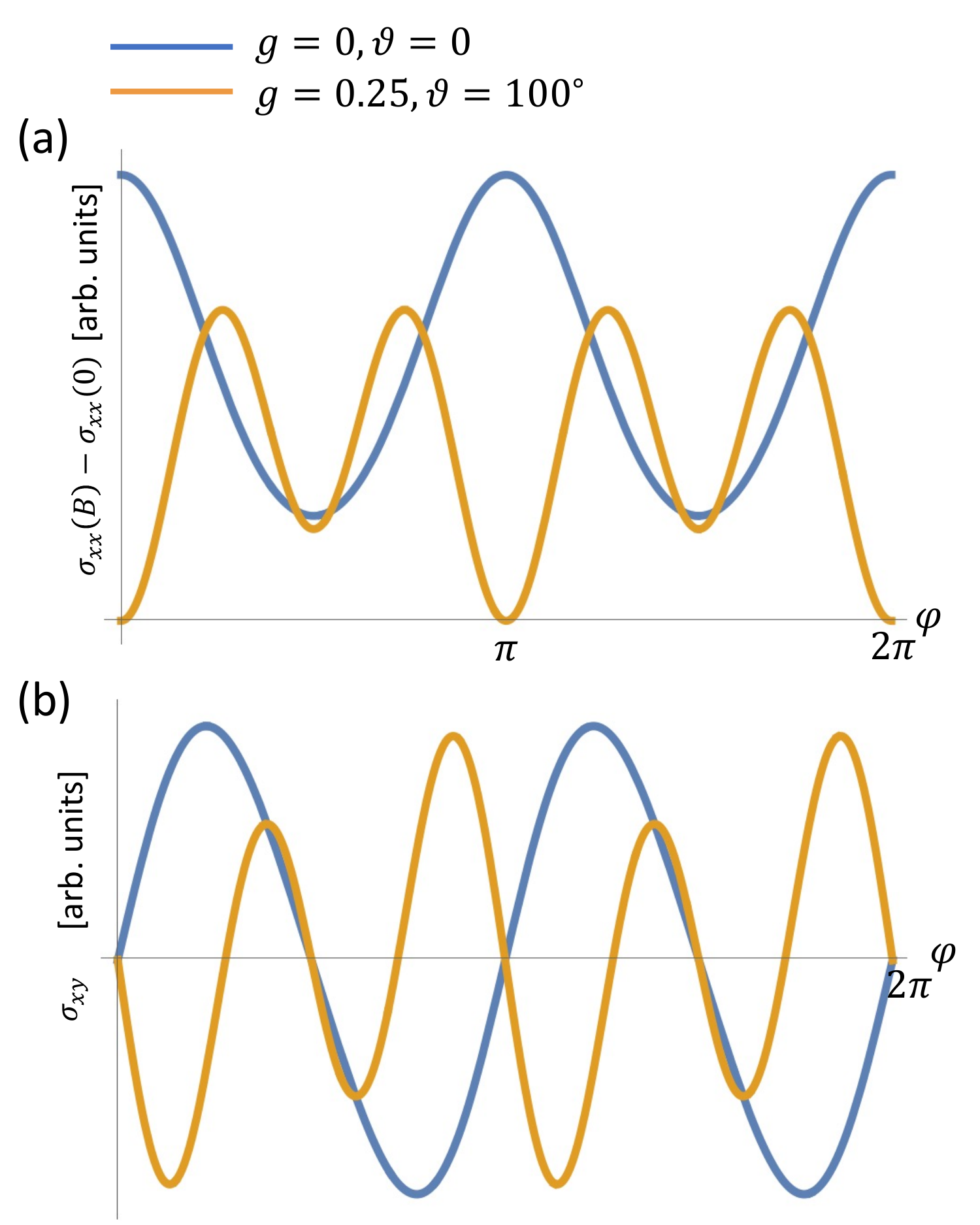}
  \caption{\label{Fig:Transport} 
  Two-fold and four-fold period in the (a) LMC and (b) PHC for the $E_{1g}$ configuration based on appropriate choice of band parameters. We have chosen $t/v = 0.2$, $\mu = 20$ meV, $v=2$ meV, and $B = 10 T$ for both cases. For period the case with finite period four component (orange), we have chosen $g = 0.25,\vartheta = 100^{\circ}$ and for the case with only period two component (blue) , we have chosen $g = 0,\, \vartheta = 0^{\circ} $.  
  }
\end{figure}

For isotropic Weyl nodes, the movement of the nodes in $\vec{k}$ space does not contribute any additional angular dependence and we obtain the familiar $\cos (2\varphi)$ and $\sin (2\varphi)$ angular dependence in LMC and PHC respectively~\cite{Nandy2017PHE}.  
Moreover, for the simple tilted Weyl nodes considered here the coefficients $a^{\chi}_{n,m}$ and $b^{\chi}_{n,m}$ are such that only even $n$ components are non-zero in the conductivity expressions in Eq.~\ref{Eq:Conductivity_total}. 
If the real space spin chirality is switched to $B_{1g}$ or $B_{2g}$ as shown in Fig.~\ref{fig:Mn3Ge_mag_cell} (b), (c), the structure gains an additional $C_{3z}$ rotation symmetry, while $I,\, M_y,\, M_z T$, and $M_xu_{1/2}T$ are still intact. 
As a result there are total $24$ iso-energy Weyl nodes(where the additional nodes are due to $C_{3z}$ symmetry). 
Following the previous analysis, adding the contribution of all these Weyl nodes and taking into account movement of the nodes, we obtain
\begin{subequations}\label{Eq:Cond_bg}
\begin{align}
    & \sigma_{xx} = 24 a^+_{0,0} + 24 a^+_{2,0} \cos (2\varphi)\, , \\
    & \sigma_{xy} = 24 b^+_{2,0} \sin (2\varphi)\, .
\end{align}
\end{subequations}

Notice that, while the expressions of the coefficients $a^{\chi}_{n,m}$ and $b^{\chi}_{n,m}$ as presented in Eqs.~\ref{Eq:App_coeff_xx},~\ref{Eq:App_coeff_xy} are specific to the tilted Weyl node model considered here, the general conductivity expressions in Eqs.~\ref{Eq:Conductivity_total},~\ref{Eq:Cond_tota_coeff}, and~\ref{Eq:Cond_bg} are independent of this detail and are obtained entirely from symmetries. 

Now we are in position to discuss possible transport signatures that can be related to the chiral order that sets up at temperatures T around 200~K. 
A period-two LMC is expected in conventional ferromagnetic metal. 
Since above $200\,\text{K}$ no long range chiral order is established, it is possible that in this regime the LMC has non-topological origin due to a finite magnetization~\cite{mcguire1975anisotropic}. 
However, any remnant zero-field magnetic moment is very small. 
The other possibility is that in the absence of long-range chiral order, the combined effect of both chiralities leads to a dominant period-two angular dependence of LMC and PHC. 
As the temperature is decreased and $E_{1g}$ chiral order is set below T$\approx200$~K, the underlying $\vec{k}\cdot\vec{p}$ expansion near iso-energy Weyl nodes is set by band parameters such that higher period LMC and PHC becomes dominant.  
In the $C_{3z}$-symmetry broken $E_{1g}$ phase, along with the change in period, the phase of the LMC and PHC signal can also change depending on the tilt angle $\vartheta$ and other band parameters. This can be seen in the Fig.~\ref{Fig:Transport}, where for the case with finite period four component (orange curves), we have chosen parameters such that the LMC has a  minima when the field is parallel to current. 
Thus maxima of the LMC need not be at $\varphi=0$ (when current and magnetic field are in same direction) as expected from chiral anomaly.  
It is important to note that although in our Eq.~(\ref{Eq:Conductivity_total}) there are finite period-six contribution, because $t^4/v^4 \ll 1$, we do not find dominant period-six LMC and PHC signal for any choice of band parameters. 
Thus our proposed increased period to four is only meant to point to a possible mechanism, since we assumed a simplified Weyl node model with elliptical Fermi surface from a single tilt vector. 
In a real system, the relevant Fermi surface around each Weyl node can be more complicated.

\section{Conclusion\label{Sec:Disc}}
In conclusion, we have sought to elucidate the connection between magnetic order and magnetotransport in Mn$_3$Ge. We performed detailed atomistic simulations of the temperature dependent magnetic order of and found a N{\'e}el temperature $T_N\approx 365$ K, which is consistent with experiments.  
The frustrated $120^{\circ}$ AFM configuration found in this system has different competing magnetic orders that are allowed by symmetries and are distinguished by the sense of rotation of localized spins over an elementary triangular plaquette (chirality). 
We find that for high temperatures, but well below $T_N$, thermal fluctuations suppress long-range chiral order. 
At a lower temperature ($\sim 200 $ K) long range chiral order is established. 
We also studied the effect of magnetic field in the $ab$-plane on the local magnetic moments. We found that for defect-free systems even in high magnetic field, the magnetic moments 
do not rotate with the magnetic field. 
However, for a system with a small defect concentration, the magnetic field can rotate moments on each site.

Based on the results of the atomistic simulations, we discuss possible distinction between transport signatures associated with different chiral orders. Using a simple tilted Weyl node model along with the crystalline symmetries of different chiral orders, we show that a chiral anomaly-induced anisotropic LMC and PHC for the $E_{1g}$ magnetic order can acquire angular dependencies [$\cos(4\varphi)$ and $\cos (6\varphi)$ terms ] that are different from the previously obtained one~\cite{Nandy2017PHE,Ma2019PHE}, while the $B_{1g}$ and $B_{2g}$ orders do not acquire these additional angular dependencies. 
In our transport analysis, the additional angular dependence comes from the motion of the Weyl nodes in the $\vec{k}$-space, which is guaranteed from spin rotation and crystal symmetries. 

Anisotropic LMC and PHC can also originate purely from the orbital effects of extremely anisotropic Fermi surfaces~\cite{Hartman1969,Collaudin2015,Zhu2018}, where the exact angular dependence is set by the type of lattice and the microscopic details of the Fermi surface. 
Although this orbital origin requires extreme anisotropy of the Fermi surface, it cannot be ruled out entirely. 
In fact the additional angular dependency acquired by the movement of the Fermi-pocket in the $\vec{k}$-space can appear in such cases as well. 
Deeper understanding of the origin of anisotropic LMC and PHC in these materials is an interesting future direction to pursue and important for potential applications that exploit the AHE or spin injection in heterostructures. 
Progress in this direction can potentially be achieved by experiments that study in detail the evolution of the amplitude of angular LMC and PHC with magnetic field and chemical potential via different doping or by gating thin films.

\section{Acknowledgement}
We are thankful to Deshun Hong and Anand Bhattacharya for numerous discussions and sharing their unpublished results. 
This work was supported the Center for the Advancement of Topological Semimetals, an Energy Frontier Research Center funded by the U.S. Department of Energy Office of Science, Office of Basic Energy Sciences. 
Research at Perimeter Institute is supported in part by the Government of Canada through the Department of Innovation, Science and Economic Development and by the Province of Ontario through the Ministry
of Economic Development, Job Creation and Trade.
We gratefully acknowledge the computing resources provided on Bebop and Blues,  high-performance computing clusters operated by the Laboratory Computing Resource Center at Argonne National Laboratory.

\appendix
\section{Derivation of LMC and PHC\label{App:LMC_PHC_derivation}}
In this section we derive the expressions for LMC and PHC used in the main text. 
Under the semiclassical Boltzmann theory, the contribution to the current density in the vicinity of a Weyl node is
\begin{align}\label{Eq:app_Jsc}
    & \vec{J}_{\chi} = -e\int \frac{d^3k}{(2\pi)^3} D^{-1}_{\chi}\dot{\vec{r}}_{\chi}f^{\chi}_{\vec{k}}(\vec{r})\, ,
\end{align}
where $\dot{\vec{r}}_{\chi}$ is the group velocity of the semiclassical wave packet, $f^{\chi}_\vec{k}(\vec{r})$ is the distribution function of the electron under applied weak external field, and $D_{\chi}(\vec{B},\vec{\Omega}^{\chi}_{\vec{k}}) = [1+(e/\hbar)(\vec{B}\cdot\vec{\Omega}^{\chi}_{\vec{k}})]^{-1}$ is the phase space volume factor and  
\begin{align}\label{Eq:app_BC}
     \vec{\Omega}^{\chi}_{\vec{k}} = -\chi \frac{ \vec{k}}{2|\vec{k}|^3}\, ,
\end{align}
is the momentum space Berry-curvature in the conduction band $\tau = +$. Since our derivation is at the fixed Fermi level, where only one of the band will have finite contribution, it suffices to the fix $\tau = +$ by choosing to work in the conduction band. 
Under the external electric field $\vec{E}$, the semiclassical wave packet dynamics leads to the equations of motion:
\begin{subequations}\label{Eq:app_EOM}
\begin{align}
    & \dot{\vec{r}}_{\chi} = \frac{1}{\hbar}\vec{\nabla}_{\vec{k}}\epsilon^{\chi}_{\vec{k}} - \dot{\vec{k}}_{\chi}\times\vec{\Omega}^{\chi}_{\vec{k}}\, ,\label{Eq:EOM1}\\
    & \dot{\vec{k}}_{\chi} = -\frac{e}{\hbar} \vec{E} - \frac{e}{\hbar}  \dot{\vec{r}}_{\chi}\times\vec{B}\, . \label{Eq:EOM2}
\end{align}
\end{subequations}

Substituting Eq.~\ref{Eq:EOM2} in Eq.~\ref{Eq:EOM1}, we obtain
\begin{align}\label{Eq:app_velocity}
    & \dot{\vec{r}}_{\chi} = D_{\chi}(\vec{B},\vec{\Omega}_{\chi}) \biggl [ \vec{v}^{\chi}_{\vec{k}} +  \frac{e}{\hbar} \vec{E} \times \vec{\Omega}^{\chi}_{\vec{k}} + \frac{e}{\hbar} (\vec{v}^{\chi}_{\vec{k}} \cdot \vec{\Omega}^{\chi}_{\vec{k}}) \vec{B} \notag\\
    &\hspace{2cm}- \frac{e}{\hbar} \vec{\Omega}^{\chi}_{\vec{k}}\cdot (\dot{\vec{k}}_{\chi}\times \vec{\Omega}^{\chi}_{\vec{k}}) \vec{B}\biggr ]\, ,
\end{align}
where 
\begin{align}\label{Eq:app_velocityFermi}
    & \vec{v}^{\chi}_{\vec{k}} = \frac{1}{\hbar} \vec{\nabla}_{\vec{k}}\epsilon^{\chi}_{\vec{k}} = \vec{t} + v\hat{\text{k}} + 4g v\frac{e}{\hbar}(\vec{B}\cdot\vec{\Omega}^{\chi}_{\vec{k}})\hat{\text{k}} +  \frac{\chi g e v}{2\hbar |\vec{k}|^2} \vec{B} 
\end{align}
is the Fermi velocity.
Up to linear order, we can ignore the last term on the RHS of Eq.~\ref{Eq:app_velocity}. 

The distribution function is calculated from the Boltzmann equation under steady state and relaxation-time approximation, i.e. $\dot{\vec{k}}_{\chi}\cdot \partial_{\vec{k}}f^{\chi}_{\vec{k}}(\vec{r}) = (f^{\chi}_{\text{eq}}-f^{\chi}_{\vec{k}}(\vec{r}))/\tau_{\vec{k}} $. 
The distribution function then can be obtained as
\begin{align}\label{Eq:app_distribution}
    & f^{\chi}_{\vec{k}} = f^{\chi}_{\text{eq}} + \biggl [ eD^{\chi}\tau_{\vec{k}} \vec{E}\cdot \vec{v}^{\chi}_\vec{k} + \frac{e^2}{\hbar} D^{\chi}\tau_{\vec{k}}(\vec{B}\cdot\vec{E})(\vec{v}^{\chi}_{\vec{k}}\cdot\Omega^{\vec{\chi}}_{\vec{k}})\notag\\
    &\hspace{2cm}+ \hat{v}^{\chi}_{\vec{k}}\cdot \Gamma^{\chi}) \biggr ]\frac{\partial f^{\chi}_{\text{eq}}}{\partial \epsilon^{\chi} }\, .
\end{align}
Here the third term on RHS is a correction factor due to magnetic field $\vec{B}$. The correction factor $\Gamma$ follows complicated coupled equations, which are obtained in Ref.~\cite{Nandy2017PHE}. However, Ref.~\cite{Ma2019PHE} shows its contribution to conductivity up to $2^{\text{nd}}$ order in magnetic field vanishes. Hence, from now on we ignore the correction factor $\Gamma$.

After substituting Eq.~\ref{Eq:app_velocity} and Eq.~\ref{Eq:app_distribution} in the current density Eq.~\ref{Eq:app_Jsc}, we obtain
\begin{widetext}
\begin{align}\label{Eq:app_Jsc2}
    & \vec{J}_{\chi} = -\frac{e^2\tau}{(2\pi)^3} \int d^3k \frac{\partial f^{\chi}_{\text{eq}}}{\partial \epsilon^{\chi} } D^{\chi} \biggl [ \vec{v}^{\chi}_{\vec{k}} + \frac{e}{\hbar} (\vec{v}^{\chi}_{\vec{k}}\cdot \Omega^{\chi}_{\vec{k}}) \vec{B}\biggr ]\biggl [ \vec{v}^{\chi}_{\vec{k}} + \frac{e}{\hbar} (\vec{v}^{\chi}_{\vec{k}}\cdot \Omega^{\chi}_{\vec{k}}) \vec{B}\biggr ]\cdot \vec{E} \notag\\
    &\hspace{4cm}-\frac{e^2\tau}{(2\pi)^3} \int d^3k \frac{\partial f^{\chi}_{\text{eq}}}{\partial \epsilon^{\chi} } D^{\chi} \frac{e}{\hbar} (\vec{E}\times \vec{\Omega}^{\chi}_{\vec{k}})\biggl [ \vec{v}^{\chi}_{\vec{k}} + \frac{e}{\hbar} (\vec{v}^{\chi}_{\vec{k}}\cdot \Omega^{\chi}_{\vec{k}}) \vec{B}\biggr ]\cdot \vec{E}\, \notag\\
    &\hspace{8cm}-\frac{e^2}{\hbar (2\pi)^3} \int d^3k  (\vec{E}\times\Omega^{\chi}_{\vec{k}}) f^{\chi}_{\text{eq}} \, .
\end{align}
\end{widetext}

Here, we have assumed that scattering rate is independent of momentum. 
The second line in the above expression is $O(E^2)$, which we ignore in the linear theory discussed here. 
The last line above contains the contribution from the occupied states, which is only finite for the transverse current. In that case, it corresponds to the anomalous Hall current. Since, it is not of interest for the discussion here, we ignore this part as well. 
From the first line, we can write down the conductivity tensor
\begin{align}\label{Eq:app_condtensor}
    & \bm{\sigma}^{\chi} =     \frac{e^2\tau}{(2\pi)^3} \int d^3k D^{\chi} \biggl [ \vec{v}^{\chi}_{\vec{k}} + \frac{e}{\hbar} (\vec{v}^{\chi}_{\vec{k}}\cdot \Omega^{\chi}_{\vec{k}}) \vec{B}\biggr ]\notag\\
    &\hspace{2cm}\biggl [ \vec{v}^{\chi}_{\vec{k}} + \frac{e}{\hbar} (\vec{v}^{\chi}_{\vec{k}}\cdot \Omega^{\chi}_{\vec{k}}) \vec{B}\biggr ] \delta(\epsilon^{\chi}_{\vec{k}}-\mu)\, .
\end{align}
Here, $\mu $ is the chemical potential. In writing down Eq.~\ref{Eq:app_condtensor}, we have taken the zero temperature limit and replaced the energy derivative of the equilibrium distribution function with the $\delta$-function. 
The final expression can be written in the form of Eq.~\ref{Eq:cond}. 
Notice that although, we have used the simple tilted Weyl node for simplicity, the expression in Eq.~\ref{Eq:cond} is more general and valid for more complicated Fermi surface surrounding the Weyl node. 
For the simple case of tilted Weyl node where the Fermi surface around the node is an ellipsoid, we can carry out the integrals in Eq.~~\ref{Eq:app_condtensor} analytically in spherical coordinates by explicitly expanding the integrand up to second order in the magnetic field. 
The final expressions for the non-zero coefficients [up to the prefactor $e^2\tau/(2\pi)^3$] are as follows:
\begin{widetext}
\begin{subequations}\label{Eq:App_coeff_xx}
\begin{align}
    & a^{\chi}_{0,0} \equiv a_{0,0} = -\frac{\pi g e^2B^2v^2}{\hbar^2} -\frac{2\pi (t^2- 2v^2)\mu^2}{3\hbar^2 v^2}  \notag\\ 
    &\hspace{1.5cm}+  \frac{\pi e^2 B^2 v^2 t^2}{210\mu^2} \biggl [\frac{14t^2}{v^2} 
    - 156 + 492 g + 115 g^2  + \frac{63 v^2}{t^2} +  \frac{371 g v^2}{t^2} +  \frac{273 g^2 v^2}{t^2} \biggr ] \\
    & a^{\chi}_{0,2} \equiv a_{0,2} = -\frac{2\pi \mu^2 t^2}{3\hbar^2 v^2} - \frac{\pi e^2 B^2 v^2 t^2}{210\mu^2} \biggl [\frac{7t^2}{v^2}    + 50 - 309 g -40 g^2 \biggr ] \\
    & a^{\chi}_{1,1} = a^{\chi}_{1,-1} = \chi a_{1,-1} = -\frac{2\pi \chi e v B t (3+g)}{3\hbar}\\
    & a^{\chi}_{2,0} \equiv a_{2,0} = -\frac{\pi g e^2B^2v^2}{\hbar^2}   \notag\\
    &\hspace{1.5cm} + \frac{\pi e^2 B^2  v^2 t^2}{420\mu^2} \biggl [\frac{7t^2}{v^2}  - 455 + 1304 g + 38 g^2  + \frac{98v^2}{t^2} +  \frac{966 g v^2}{t^2} +  \frac{98 g^2 v^2}{t^2} \biggr ] \\
    & a^{\chi}_{2,-2} \equiv a_{2,-2} = -\frac{\pi e^2 B^2 t^4 }{15\mu^2} + \frac{\pi e^2 B^2 v^2 t^2}{210 \mu^2} \biggl [ - 113 + 421 g + 40 g^2 \biggr ] \\
    & a^{\chi}_{2,-4}  \equiv a_{2,-4} = \frac{\pi e^2 t^4 B^2}{60 \mu^2} 
\end{align}
\end{subequations}
and

\begin{subequations}\label{Eq:App_coeff_xy}
\begin{align}
    & b^{\chi}_{0,2} \equiv b_{0,2} = -\frac{2\pi \mu^2 t^2}{3\hbar^2 v^2} - \frac{\pi e^2 B^2 v^2 t^2}{210\mu^2} \biggl [\frac{7t^2}{v^2}  + 50 - 309 g -40 g^2 \biggr ] \\
    & b^{\chi}_{1,-1} = \chi b_{1,-1} = \frac{4\pi \chi e v B t g}{3\hbar}\\
    & b^{\chi}_{1,1} = \chi b_{1,1} = \frac{2\pi \chi e v B t (g-3)}{3\hbar}\\
    & b^{\chi}_{2,0} \equiv b_{2,0} = -\frac{\pi g e^2B^2v^2}{\hbar^2}   \notag\\
    &\hspace{1.5cm} + \frac{\pi  e^2 B^2 v^2 t^2}{420\mu^2} \biggl [\frac{7t^2}{v^2}  - 405 + 1304 g + 38 g^2 + \frac{98v^2}{t^2} +  \frac{966 g v^2}{t^2} +  \frac{98 g^2 v^2}{t^2} \biggr ]\\
    & b^{\chi}_{2,-2} \equiv b_{2,-2} =  -\frac{\pi e^2 B^2 t^4 }{15\mu^2} + \frac{\pi e^2 B^2 v^2 t^2}{210 \mu^2} \biggl [  - 113 + 421 g + 40 g^2 \biggr ] \\
    & b^{\chi}_{2,-4} \equiv b_{2,-4} = -\frac{\pi e^2 B^2 t^4}{60\mu^2}
\end{align}
\end{subequations}
\end{widetext}



\bibliography{bibliography}

\end{document}